\documentclass{article}
\usepackage{spconf}
\usepackage[utf8]{inputenc} 
\usepackage[T1]{fontenc}    
\usepackage{hyperref}       
\usepackage{url}            
\usepackage{booktabs}       
\usepackage{amsfonts}       
\usepackage{nicefrac}       
\usepackage{microtype}      
\usepackage{lipsum}
\usepackage{fancyhdr}       
\usepackage{graphicx}       
 \graphicspath{{media/}}     
\usepackage{amsmath}
\usepackage{amssymb}
\usepackage{mathrsfs}
\usepackage{braket}
\usepackage{xcolor}
\usepackage{tikz}
\usepackage{quantikz}
\usepackage{braket}
\usepackage{mathtools}
\usepackage[nolist,nohyperlinks]{acronym}
\usepackage[acronym]{glossaries}
\setacronymstyle{long-short}
\usepackage[numbers,sort&compress]{natbib}
\usepackage{subcaption}

\usepackage{multirow}
\usepackage{enumitem}
\usepackage{xspace}
\usepackage{comment}
\newcommand{\design}{{TGen-SQD}\xspace}

\title{\design: Physics-Informed Transferable Configuration Generator \\for Sample-Based Quantum Diagonalization\vspace{-10pt}}
\name{Niloy Deb Roy Mishu$^*$ \quad Lei Jiang$^*$ \quad Mohsen Heidari$^*$ \quad Ruizhe Zhang$^\dagger$ \quad Fan Chen$^*$\vspace{-10pt}}
\address{$^\dagger$Department of Computer Science, Purdue University, West Lafayette, IN, USA\\
$^*$Luddy School of Informatics, Computing, and Engineering, Indiana University, Bloomington, IN, USA\vspace{-16pt}}

\begin{document}
%
\maketitle
\begin{abstract}
\vspace{-6pt}
Sample-based quantum diagonalization (SQD) avoids iterative variational circuit optimization, but its accuracy and cost depend on selecting a compact determinant subspace that captures the ground state. Existing machine-learning selectors use task-specific physical cues and typically retrain for each Hamiltonian. We introduce~\design, a physics-informed generator jointly trained across multiple Hamiltonians. It learns from configuration-interaction-weighted data and uses orbital- and system-level descriptors to guide generation. After training, its weights remain frozen for inference on seen and zero-shot geometries. On a 20-qubit H$_{10}$ chain, TGen-SQD achieves mean energy errors of $40.18 \pm 1.52$ and $43.97 \pm 1.20$~mHa on seen and unseen geometries, respectively—12.7\% and 11.7\% lower than the state-of-the-art PIGen-SQD baseline under matched shot and proposal budgets.
\end{abstract}


\newacro{nisq}[NISQ]{noisy-intermediate scale quantum}
\newacro{vqe}[VQE]{variational quantum eigensolver}
\newacro{pqc}[PQC]{parameterized quantum circuit}
\newacro{qsci}[QSCI]{quantum-selected configuration interaction}
\newacro{sqd}[SQD]{sample-based quantum diagonalization}
\newacro{ci}[CI]{configuration interaction}
\newacro{cr}[CR]{configuration recovery}
\newacro{hf}[HF]{Hartree-Fock}
\newacro{arnn-sci}[ARNN-SCI]{Autoregressive Neural Network Selected \ac{ci}}
\newacro{hi-nqs}[HI-NQS]{Handover Iterative Neural Quantum State}
\newacro{haar-sci}[HAAR-SCI]{Hamiltonian-guided autoregressive selected-\ac{ci}}
\newacro{mppt}[MPPT]{Møller-Plesset perturbation theory}
\newacro{rbm}[RBM]{restricted Boltzmann machine}
\newacro{cd}[CD]{contrastive divergence}
\newacro{gru}[GRU]{gated recurrent unit}

\vspace{-10pt}
\section{Introduction}
\label{sec:intro}
\vspace{-10pt}

Accurate estimation of molecular ground-state energies is essential to many open problems in chemistry~\cite{nam2020ground}, yet classical computation is limited by the combinatorial growth of the many-electron Hilbert space~\cite{mcardle2020quantum}. Near-term hybrid quantum--classical methods~\cite{bharti2022noisy} address this challenge through approaches such as the variational quantum eigensolver (VQE)~\cite{peruzzo2014variational}, the quantum approximate optimization algorithm (QAOA)~\cite{farhi2014quantum}, and quantum-selected configuration interaction (QSCI), including sample-based quantum diagonalization (SQD)~\cite{kanno2026qsci,robledo2025sqd}. Unlike VQE and QAOA, SQD samples computational-basis configurations and classically diagonalizes the Hamiltonian in their Slater-determinant subspace, avoiding iterative variational circuit optimization and its associated barren-plateau problem~\cite{wang2021noise}. However, its accuracy and cost depend on the quality and diversity of sampled determinants: missing important configurations raises the variational energy and limits recovered correlation energy, while including low-value configurations increases subspace construction, storage, and diagonalization costs~\cite{reinholdt2025limitations,vaquero2026recovery,goings2021rlci,rende2025foundation}. Thus, SQD must identify a compact, representative subspace under finite sampling and classical-computation budgets~\cite{vaquero2026recovery}.

\textbf{Limitations of Prior Methods.}
Recent machine-learning approaches~\cite{thompson2026arnnsci,zhang2025haarsci,chang2026hinqs} improve configuration selection through learned sampling and physics-guided screening. PIGen-SQD~\cite{patra2026pigensqd} further enriches quantum samples with \ac{mppt}-selected determinants to seed an \ac{rbm} generator. However, these approaches leave two limitations:
\begin{itemize}[leftmargin=*, topsep=0pt, partopsep=0pt, itemsep=0pt]
    \item Existing methods encode physical information through task-specific cues—Hamiltonian-coupling scores, perturbative rankings, or seed-pool enrichment—rather than directly incorporating a broad set of descriptors, including orbital energies, one- and two-electron integrals, occupations, geometry, and selected-subspace information, as explicit context for a generator of configuration-interaction (CI) determinants.
    
    \item Generators are generally trained or updated separately for each target Hamiltonian, limiting transfer across related geometries and adding retraining cost~\cite{coe2019curves,goings2021rlci,rende2025foundation}. For example, PIGen-SQD~\cite{patra2026pigensqd} retrains its \ac{rbm} via \ac{cd} for each target Hamiltonian.
\end{itemize}

\textbf{Contributions}.
In this work, we propose~\design, a physics-informed, transferable configuration generator for SQD, and make the following contributions:
\begin{itemize}[leftmargin=*, topsep=-2pt, partopsep=-2pt, itemsep=-2pt]
    \item \textbf{Physics-informed configuration representation.}
     We form configuration-interaction-weighted training data from local unitary cluster--Jastrow (LUCJ) samples, the Hartree--Fock (HF) determinant, and rank-three MPPT anchors. Feature-specific tokenization encodes orbital- and system-level context—including energies, one- and two-electron integrals, occupations, geometry, and selected-subspace properties—for conditioning the generator.

    \item \textbf{Transferable autoregressive generation.}
     We train a gated recurrent unit (GRU) based autoregressive generator by maximum likelihood on CI-weighted data from nine Hamiltonians. At inference, its weights remain frozen across seen (in-distribution) and unseen (zero-shot) geometries; only the target-specific context and selected-CI subspace are updated. This enables target-conditioned configuration generation without per-Hamiltonian retraining or Gibbs sampling.

    \item \textbf{Improved accuracy under matched sampling budgets.}
     Our experiments on a $20$-qubit $H_{10}$ chain show that TGen-SQD, jointly trained across nine Hamiltonians, trains stably and consistently outperforms PIGen-SQD~\cite{patra2026pigensqd} on seen and zero-shot (unseen) geometries as the proposal budget increases. At matched target-shot and proposal budgets, TGen-SQD achieves mean energy errors of $40.18 \pm 1.52$ and $43.97 \pm 1.20$~mHa on seen and unseen geometries, respectively, corresponding to reductions of $12.7\%$ and $11.7\%$ relative to PIGen-SQD.
\end{itemize}

\vspace{-12pt}
\section{Background and Motivation}
\label{sec:background}

\vspace{-6pt}
\subsection{SQD Workflow}
\vspace{-4pt}

SQD combines quantum sampling with classical subspace diagonalization. At a fixed molecular geometry, each bitstring $x$ \underline{\textit{encodes a Slater determinant}} $\lvert\Phi_x\rangle$, with 1 and 0 denoting occupied and unoccupied spin orbitals. 
A local unitary cluster Jastrow (LUCJ) circuit \cite{motta2023lucj} \underline{\textit{prepares a state}} from the Hartree--Fock (HF) reference: $\lvert\psi_q\rangle$$=$$U_{\mathrm{LUCJ}}(\theta)\lvert\Phi_{\mathrm{HF}}\rangle$, and computational-basis measurements sample determinants with probabilities $p_q(x)=\lvert\langle\Phi_x\vert\psi_q\rangle\rvert^2$ \cite{robledo2025sqd}. 

To \underline{\textit{augment the sampled configurations}}, many-particle perturbation theory (MPPT) identifies determinants with large perturbative contributions. The wavefunction is expressed as a sum of corrections to the HF reference,
\vspace{-4pt}
\begin{equation}
    \lvert\Psi_0\rangle \propto
    \lvert\Phi_{\mathrm{HF}}\rangle
    + \lvert\Psi^{(1)}\rangle
    + \lvert\Psi^{(2)}\rangle + \cdots,
    \label{eq:mppt-expansion}
    \vspace{-4pt}
\end{equation}
where the superscripts indicate the correction order. The first-order perturbative correction contains double excitations,
\vspace{-4pt}
\begin{equation}
    \lvert\Psi^{(1)}\rangle
    = \sum_{\substack{i<j\\a<b}}
    t_{ij}^{ab}\lvert\Phi_{ij}^{ab}\rangle,
    \qquad
    t_{ij}^{ab}
    = \frac{\langle ab\Vert ij\rangle}
    {\epsilon_i+\epsilon_j-\epsilon_a-\epsilon_b}.
    \label{eq:mppt-doubles}
    \vspace{-6pt}
\end{equation}
where $\lvert\Phi_{ij}^{ab}\rangle$ is formed by exciting electrons from occupied HF spin orbitals $i,j$ to virtual orbitals $a,b$. Here, $\epsilon_p$ is an orbital energy, and $\langle ab\Vert ij\rangle$ is an antisymmetrized two-electron integral~\cite{moller1934perturbation}. Double excitations with large $|t_{ij}^{ab}|$ are retained; at rank three, second-order terms generate additional single and triple excitations through sparse operations on lower-rank amplitudes and integrals~\cite{patra2026pigensqd}.

The sampled and perturbatively selected determinants define a subspace $\mathcal{S}$. The Hamiltonian is constructed and \underline{\textit{classically diagonalized}} in this subspace, for example using the Davidson algorithm~\cite{davidson1975}:
\vspace{-4pt}
\begin{equation}
    (H_{\mathcal{S}})_{xy}
    = \langle\Phi_x\vert\hat{H}\vert\Phi_y\rangle,
    \qquad
    H_{\mathcal{S}}\mathbf{c}
    = E_{\mathcal{S}}\mathbf{c}.
    \label{eq:selected-ci}
    \vspace{-4pt}
\end{equation}
The lowest eigenpair $(E_{\mathcal{S}},\mathbf{c})$ yields the energy estimate
$E_{\mathcal{S}}$ and the selected-CI state
$\lvert\Psi_{\mathcal{S}}\rangle=\sum_{x\in\mathcal{S}}c_x\lvert\Phi_x\rangle$
where $c_x$ is the component of the lowest-eigenvalue eigenvector
associated with determinant $\lvert\Phi_x\rangle$.

\vspace{-4pt}
\subsection{Related Works and Motivation}
\vspace{-4pt}

Table~\ref{tab:method-comparison} compares recent machine-learning approaches to configuration selection. 
ARNN-SCI~\cite{thompson2026arnnsci} learns from successive CI solutions; 
HAAR-SCI~\cite{zhang2025haarsci} uses Hamiltonian-coupling screening; 
HI-NQS~\cite{chang2026hinqs} combines spin-aware generation with perturbative ranking; and 
PIGen-SQD~\cite{patra2026pigensqd} seeds an RBM with MPPT-selected determinants and generates candidates by Gibbs sampling. 
These methods use task-specific physical cues, and none reuses a generator across Hamiltonians. 
These gaps motivate us to investigate transfer learning for SQD through~\design, which combines MPPT-enriched training data and orbital- and system-level context in a jointly trained generator reused across related Hamiltonians.

\begin{table}[t]
    \centering
    \caption{Comparison between~\design~and previous works ($\checkmark$ = full support, $\triangle$ = partial support, --- = not supported.).}
    \label{tab:method-comparison}
    \vspace{-10pt}
    \footnotesize
    \setlength{\tabcolsep}{1pt}
    \renewcommand{\arraystretch}{0.95}

    \begin{tabular*}{\columnwidth}
        {@{\extracolsep{\fill}}lcccc@{}}
        \toprule
        Method
        & \shortstack{Autoregressive\\sampling}
        & \shortstack{Physics-\\informed}
        & \shortstack{MPPT-enriched\\seed pool}
        & \shortstack{Transferability} \\
        \midrule
        ARNN-SCI~\cite{thompson2026arnnsci}
        & $\checkmark$ & --- & --- & --- \\
        HAAR-SCI~\cite{zhang2025haarsci}
        & $\checkmark$ & $\triangle$ & --- & --- \\
        HI-NQS~\cite{chang2026hinqs}
        & $\checkmark$ & $\triangle$ & --- & --- \\
        PIGen-SQD~\cite{patra2026pigensqd}
        & --- & $\triangle$ & $\checkmark$ & --- \\
        \textbf{\design}
        & $\checkmark$ & $\checkmark$ & $\checkmark$ & $\checkmark$ \\
        \bottomrule
    \end{tabular*}
    \vspace{-18pt}
\end{table}

\vspace{-12pt}
\section{\design~Framework}
\label{sec:method}
\vspace{-6pt}

\raggedbottom
\begin{figure*}[!t]
    \centering
    \includegraphics[width=\textwidth]{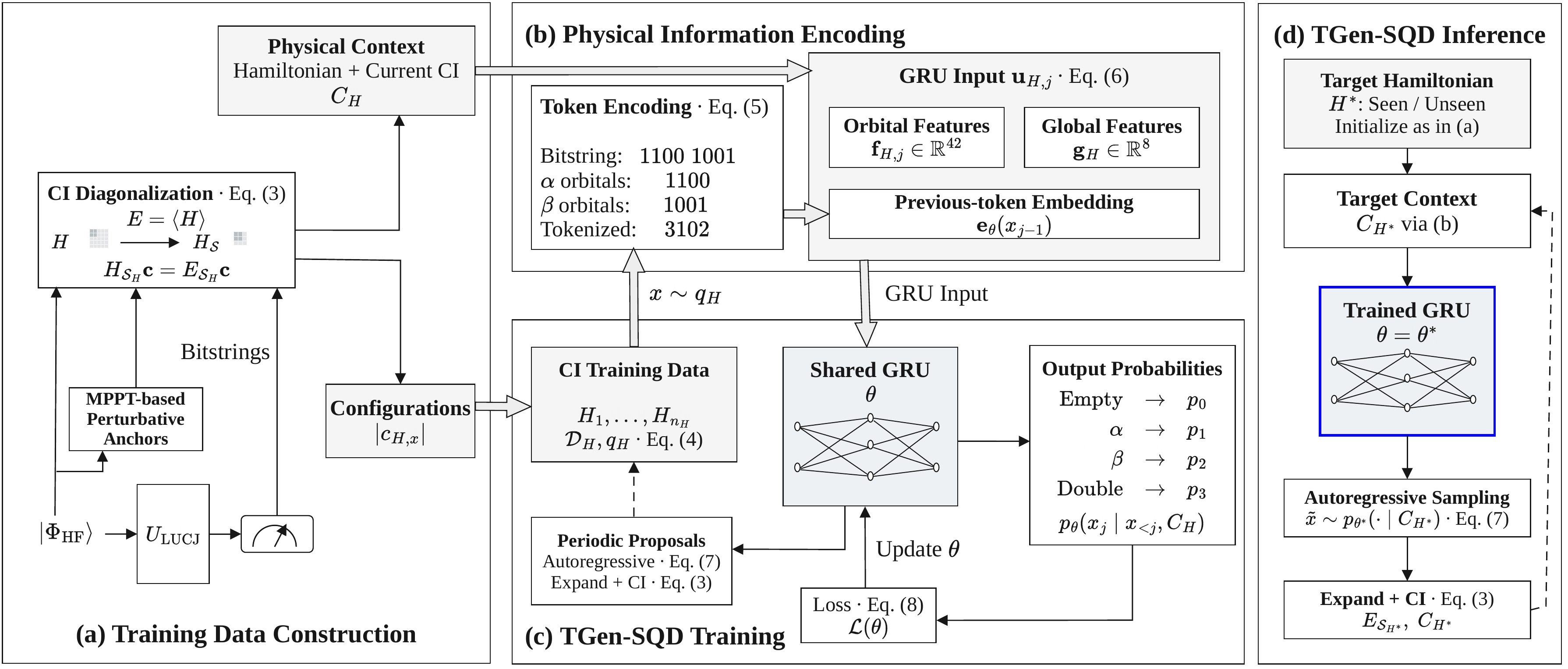}
    \vspace{-18pt}
    \caption{Proposed~\design~framework: (a) training-data construction from LUCJ samples, HF, and perturbative anchors; (b) physics-informed encoding via configuration tokenization and orbital and global feature embeddings; (c) joint training with periodic CI expansion; and (d) in-distribution evaluation on seen and zero-shot inference on unseen geometries.}    
    \label{fig:framework}
    \vspace{-14pt}
\end{figure*}

Fig.~\ref{fig:framework} overviews the propose~\design~workflow: physics-informed configurations and Hamiltonian descriptors are encoded for joint training of a shared GRU, whose weights are then frozen for inference on related Hamiltonians.

\vspace{-10pt}
\subsection{Training Data Construction}
\label{subsec:physics-data}
\vspace{-4pt}

Let $\{H_i\}^{n_H}_{i=1}$ denote the training Hamiltonians. For each Hamiltonian $H$, we use the LUCJ sampling and perturbative enrichment described in Sec.~\ref{sec:background} \cite{patra2026pigensqd}. All unique configurations from LUCJ measurements are combined with the \ac{hf} determinant and perturbative anchors up to excitation rank three. After adding configurations and removing duplicates, we obtain the selected space $\mathcal{S}_H$. Diagonalization using \eqref{eq:selected-ci} gives its energy and coefficients $c_{H,x}$ (Fig.~\ref{fig:framework}(a)).

Each Hamiltonian maintains its own training dataset, $\mathcal{D}_H$. $\mathcal{D}_H$ excludes the \ac{hf} determinant and configurations with $|c_{H,x}|\leq 10^{-9}$. Within each Hamiltonian task, configurations are sampled according to
\vspace{-4pt}
\begin{equation}
    q_H(x)=\frac{|c_{H,x}|}
    {\sum_{y\in\mathcal{D}_H}|c_{H,y}|},
    \qquad x\in\mathcal{D}_H.
    \label{eq:ci-training-weights}
    \vspace{-4pt}
\end{equation}

\ac{ci} training data for our model combines individual Hamiltonian datasets.

\vspace{-10pt}
\subsection{Physical Information Encoding}
\label{subsec:encoding}
\vspace{-4pt}

\noindent\textbf{Configuration Tokenization.}
For a system with $M$ spatial orbitals, each configuration $x$ comprises $2M$ occupation bits. We pair the $\alpha$- and $\beta$-spin bits for each orbital $j$ and map them to one of four tokens using Eq.~\ref{eq:occupation-token}, where $n_j^\alpha,n_j^\beta\in\{0,1\}$: empty, $\alpha$-only, $\beta$-only, or doubly occupied. Thus, each configuration maps to an $M$-token sequence. For example, the bitstring $1100\,1001$ in Fig.~\ref{fig:framework}(b) has $\alpha$- and $\beta$-occupations $1100$ and $1001$, respectively, yielding the token sequence $3102$ (double, $\alpha$-only, empty, $\beta$-only).
\vspace{-8pt}
\begin{equation}
    x_j=n_j^{\alpha}+2n_j^{\beta}\in\{0,1,2,3\},
    \label{eq:occupation-token}
    \vspace{-8pt}
\end{equation}

\noindent\textbf{Physical Context Embedding.}
Orbital features $\mathbf f_{H,j}$ include orbital energies, one-electron and Coulomb/exchange matrix features, and \ac{hf}/current-\ac{ci} occupations. Global features $\mathbf g_H$ describe system size, electron filling, geometry, reference and correlation energies, and selected-space size. Together, these features form the task context $C_H$.

\noindent\textbf{GRU Input Representation.}
To predict $x_j$, we concatenate a one-hot encoding or learned embedding of the previous token with the current-orbital and global features, as defined in Eq.~\ref{eq:gru-input}. Here, semicolons denote concatenation, and $x_0$ is a special start token. The previous token is observed during training and sampled during inference.
\vspace{-8pt}
\begin{equation}
    \mathbf u_{H,j}=
    [\mathbf e_{\theta}(x_{j-1});\mathbf f_{H,j};\mathbf g_H],
    \label{eq:gru-input}
    \vspace{-8pt}
\end{equation}

\vspace{-6pt}
\subsection{Joint Training Across Multiple Hamiltonians}
\label{subsec:shared-training}
\vspace{-6pt}

All Hamiltonians share an embedding layer, \ac{gru}, and output head with parameters $\theta$. At each step, the recurrent state summarizes preceding tokens, and the output head produces four logits. We mask occupations that would make the required $\alpha$- and $\beta$-electron counts unattainable, then apply softmax to obtain probabilities over the four occupation states of $x_j$. The resulting autoregressive distribution is given in Eq.~\ref{eq:autoregressive-distribution}, where $x_{<j}$ denotes the preceding tokens.

\vspace{-8pt}
\begin{equation}
    p_\theta(x\mid C_H)=
    \prod_{j=1}^{M}p_\theta(x_j\mid x_{<j},C_H),
    \label{eq:autoregressive-distribution}
    \vspace{-8pt}
\end{equation}

\vspace{-4pt}
\begin{equation}
    \mathcal L(\theta)=
    -\frac{1}{n_H}\sum_{i=1}^{n_H}
    \mathbb E_{x\sim q_{H_i}}
    \left[\frac{1}{M}\log p_\theta(x\mid C_{H_i})\right].
    \label{eq:shared-training-loss}
    \vspace{-4pt}
\end{equation}

We provide observed previous tokens as inputs and maximize the likelihood of the target tokens using minibatches balanced across Hamiltonians. Equivalently, we minimize the negative log-likelihood in Eq.~\ref{eq:shared-training-loss}, where $1/n_H$ and $1/M$ average over Hamiltonians and orbital tokens, respectively. Periodically between training updates, the generator proposes configurations for each training Hamiltonian. We add them to $\mathcal S_H$, rediagonalize the subspace Hamiltonian, and update $\mathcal D_H$, $q_H$, and $C_H$. After training, we freeze the generator and retain the learned weights $\theta^*$ for inference.

\vspace{-6pt}
\subsection{In-Distribution and Zero-Shot Inference}
\label{subsec:frozen-inference}
\vspace{-6pt}

For each target Hamiltonian $H$, seen or unseen during training, we initialize a fresh selected space $\mathcal S_H$ and context $C_H$ as in Sec.~\ref{subsec:physics-data}. Conditioned on $C_H$, we generate configurations one token at a time, starting from $x_0$ and feeding each sampled token back as the next input. Configurations are sampled autoregressively using the target Hamiltonian and current-CI context. The token sequences are decoded into $2M$-bit configurations. Newly sampled unique configuration proposals and their spin mirrors are merged into $\mathcal S_H$. We re-diagonalize the enlarged space using \eqref{eq:selected-ci} and update its energy and the \ac{ci}-derived physics features in $C_H$. This cycle repeats under a fixed proposal budget, and returns the final selected-\ac{ci} energy $E_{\mathcal S_H}$. The same trained weights $\theta^*$ remain fixed across all rounds and targets.

\vspace{-10pt}
\section{Experimental Setup and Results}
\label{sec:experiments}
\vspace{-8pt}

\subsection{Experimental Setup}
\label{subsec:experimental_setup}
\vspace{-4pt}

\textbf{Benchmarks}.
We study linear H$_{10}$ chains in the STO-3G basis with $20$ qubits ($2M$), $10$ spatial orbitals ($M$), and $10$ electrons ($N_\alpha=N_\beta=5$). We train on nine Hamiltonians with bond lengths from $1.10$ to $1.50$~\AA{} in $0.05$~\AA{} increments. The validation, seen-test, and unseen-test sets each contain four geometries, spanning $1.125$--$1.425$, $1.15$--$1.45$, and $1.175$--$1.475$~\AA{}, respectively, in $0.10$~\AA{} increments. The PySCF library supplies \ac{hf}, CCSD, selected-CI, and FCI calculations; ffsim simulates two-repetition, CCSD-initialized LUCJ states with $256$ ideal shots per Hamiltonian. All generators receive the same rank-three, perturbatively enriched initial space.

\begin{figure}[!t]
    \centering
    \includegraphics[width=0.5\linewidth]{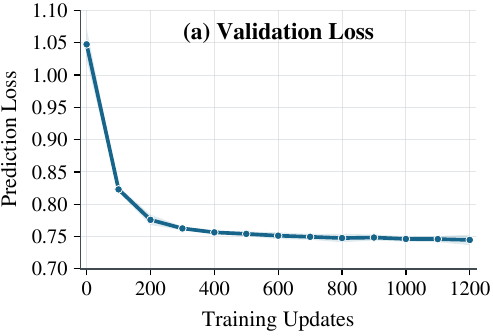}%
    \includegraphics[width=0.5\linewidth]{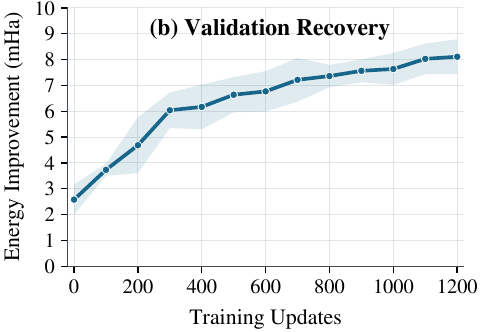}
    \vspace{-18pt}
    \caption{\design~validation metrics over training: (a) prediction loss and (b) energy improvement.}
    \label{fig:validation-results}
    \vspace{-12pt}
\end{figure}

\textbf{Simulation}.
We implement~\design~in PyTorch using learned $16$-dimensional token embeddings, $42$ orbital features, $8$ global features, and a single GRU layer with $64$ hidden units. 
Our common recovery routine omits the blacklist/pruning. The local GRU receives $100$ updates per Hamiltonian per recovery round. Frozen shared-GRU sampling uses temperature $T=2$. Each generator receives six rounds of $1{,}270$ configuration proposals per target, totaling $7{,}620$ proposals, including duplicates. 
We train for $1{,}200$ Adam updates at a learning rate of $5\mathrm{e}{-4}$ with balanced minibatches of size $128$, expanding the configuration space every $200$ updates. Checkpoints are evaluated every $100$ updates using $2{,}540$ proposals per validation geometry, and the one with the highest validation energy improvement is selected.

\textbf{Evaluation Metrics and Schemes}.
Following prior work~\cite{thompson2026arnnsci, chang2026hinqs, patra2026pigensqd}, we measure performance by the energy error (Eq.~\ref{eq:energy-error}) between the selected-configuration energy $E_{\mathcal S_H}$ and reference ground-state energy $E_H^{\mathrm{FCI}}$.
For each seed, we average the energy error over the four geometries, then report the mean and $95\%$ confidence interval across the five seeds.
\vspace{-6pt}
\begin{equation} 
    \Delta E_H = 10^3\left(E_{\mathcal S_H}-E_H^{\mathrm{FCI}}\right) \quad \text{(mHa)}, 
    \label{eq:energy-error} 
    \vspace{-6pt}
\end{equation} 

We compare the following schemes:
\begin{itemize}[leftmargin=*, topsep=-2pt, partopsep=-2pt, itemsep=-2pt]
    \item \textit{Base}: CI energy of the physics-informed initial space, without any training or generative model.
    \item \textit{RBM (Untrained)}: Gibbs sampling from random RBM weights, reinitialized for each target and recovery round.
    \item \textit{PIGen-SQD}~\cite{patra2026pigensqd}: RBM baseline using the released routines, with CD-20 and three training epochs per round.
    \item \textit{GRU (Untrained)}:  A separate, randomly initialized GRU generates configurations for each target with fixed weights.    
    \item \textit{TGen-SQD (w/o TL)}: A separate GRU is trained from scratch on each target Hamiltonian's evolving CI data.
    \item \textit{TGen-SQD (w/ TL)}: One jointly trained GRU is reused across all targets with frozen weights and updated target contexts.
\end{itemize}

\begin{figure}[!t]
    \centering
    \includegraphics[width=0.5\linewidth]{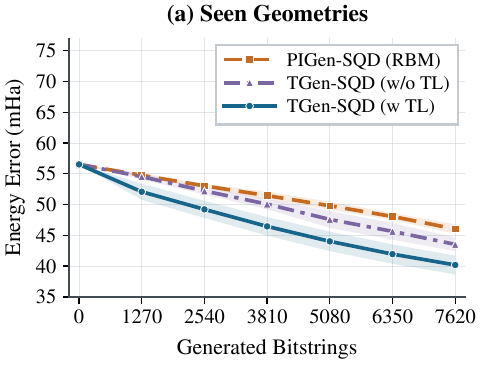}%
    \hspace{-1.5pt}%
    \includegraphics[width=0.5\linewidth]{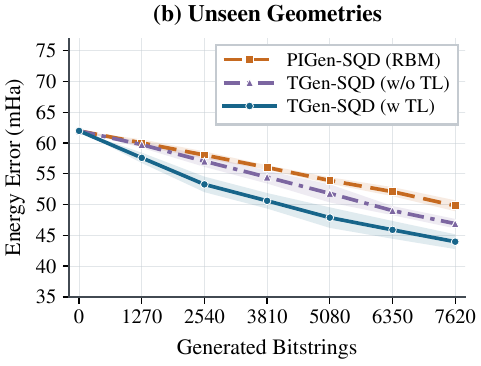}
    \vspace{-20pt}
    \caption{Performance comparison of~\design~and PIGen-SQD~\cite{patra2026pigensqd} on seen (a) and unseen (b) test geometries.}
    \label{fig:test-results}
    \vspace{-18pt}
\end{figure}
\begin{table}[!t]
    \centering
    \caption{Overall performance comparison among all schemes in energy-error (mHa) . $\Delta$ denotes the improvement of~\design~over the SOTA PIGen-SQD~\cite{patra2026pigensqd}.}    
    \label{tab:results}    
    \vspace{-10pt}
    \footnotesize
    \setlength{\tabcolsep}{1pt}
    \renewcommand{\arraystretch}{0.95}

        \begin{tabular*}{\columnwidth}
        {@{\extracolsep{\fill}}lcc@{}}
        \toprule
        Method & In-Distribution & Zero-Shot  \\
        \midrule
        \shortstack[l]{Base}
        & $56.53$ & $61.98$ \\\hline
        RBM (Untrained)
        & $49.29 \pm 0.50$ & $53.85 \pm 0.75$ \\
        PIGen-SQD (RBM)
        & $46.02 \pm 0.74$ & $49.81 \pm 0.89$ \\\hline
        GRU (Untrained)
        & $50.27 \pm 1.00$ & $54.13 \pm 1.14$ \\
        \design~(w/o TL)
        & $43.53 \pm 1.10$ & $46.90 \pm 0.77$ \\
        \textbf{\design~(w/ TL)}
        & $\mathbf{40.18}\pm\mathbf{1.52}$
        & $\mathbf{43.97}\pm\mathbf{1.20}$ \\
        \textbf{$\Delta$}
        &\textcolor{red}{\textbf{12.7\%$\downarrow$}}
        &\textcolor{red}{\textbf{11.7\%$\downarrow$}}\\
        \bottomrule
    \end{tabular*}
    \vspace{-16pt}
\end{table}

\vspace{-8pt}
\subsection{Experimental Results}
\label{subsec:results}
\vspace{-4pt}

\textbf{Validation and Test Stability}.
We track the shared GRU's validation metrics over training updates. 
Despite conditioning on orbital- and system-level physics-informed features, the prediction loss decreases from $1.047$ to $0.745$ and then levels off (Fig.~\ref{fig:validation-results}(a)), 
while validation energy improvement rises from $2.58$ to $8.11$~mHa under a fixed generation budget (Fig.~\ref{fig:validation-results}(b)). 
These trends indicate that the model trains effectively with the physics-informed features.

\textbf{Comparison with SOTA PIGen-SQD~\cite{patra2026pigensqd}}.
Fig.~\ref{fig:test-results} shows the mean energy error on seen and unseen test geometries as the cumulative number of generated bitstrings increases. After each round of $1{,}270$ proposals, the expanded selected space is diagonalized, for a total of $7{,}620$ proposals. \design, jointly trained on physics-informed, perturbatively enriched data and used with frozen weights, achieves lower mean error than PIGen-SQD and individually trained \design as the proposal budget increases on both sets of geometries. This demonstrates that the shared generator transfers across geometries without target-specific retraining.

\textbf{Overall Performance Comparison}.
Table~\ref{tab:results} compares the Base, RBM, and GRU groups. Both untrained generators improve on the physics-informed Base, showing that sampling and CI expansion recover useful configurations without neural training. Training further improves both model families, with individually trained \design~already outperforming PIGen-SQD. Shared training provides an additional advantage, achieving the lowest mean errors on both seen and unseen geometries. Under matched target-shot and configuration-proposal budgets, \design~with transfer learning reduces mean error relative to PIGen-SQD by $12.7\%$ on seen geometries and $11.7\%$ on unseen geometries. It also improves over its individually trained variant by $3.35$ and $2.93$~mHa, respectively. This demonstrates benefits from learning across Hamiltonians even for seen targets. The retained advantage on unseen Hamiltonians indicates that shared training learns transferable electron-occupation patterns. This enables \design~to recover useful configurations across targets using their physical contexts, with frozen weights and without target-specific neural retraining.

\vspace{-10pt}
\section{Conclusion}
\vspace{-4pt}
We introduced~\design, a physics-informed autoregressive generator jointly trained across related Hamiltonians and reused with frozen weights for seen and zero-shot geometries. On a 20-qubit H$_{10}$ chain, it reduced mean energy errors by 12.7\% and 11.7\% versus PIGen-SQD under matched sampling budgets, demonstrating the value of transferable, physically informed configuration generation for SQD.

\clearpage
\bibliographystyle{IEEEbib}
\bibliography{references}

\end{document}